\documentclass[letterpaper]{article} % DO NOT CHANGE THIS
\usepackage{aaai2026}                % DO NOT CHANGE THIS
\usepackage{times}                   % DO NOT CHANGE THIS
\usepackage{helvet}                  % DO NOT CHANGE THIS
\usepackage{courier}                 % DO NOT CHANGE THIS
\usepackage[hyphens]{url}            % DO NOT CHANGE THIS
\usepackage{graphicx}                % DO NOT CHANGE THIS
\usepackage{enumitem}
\usepackage{booktabs}
\usepackage{algorithm}
\usepackage{algorithmic}
\usepackage{amsmath}
\usepackage{amssymb}
\usepackage{newfloat}
\usepackage{listings}
\usepackage{natbib}                  % DO NOT ADD ANY OPTIONS TO IT
\usepackage{caption}                 % DO NOT ADD ANY OPTIONS TO IT
\usepackage{multirow}
\floatstyle{ruled}
\newfloat{listing}{tb}{lst}{}
\floatname{listing}{Listing}

\DeclareCaptionStyle{ruled}{labelfont=normalfont,labelsep=colon,strut=off}
\floatstyle{ruled}
\newfloat{listing}{tb}{lst}{}
\floatname{listing}{Listing}

\title{LLM-Aligned Geographic Item Tokenization for Local-Life Recommendation}
\author {
Hao Jiang\textsuperscript{\rm 1} \thanks{Equal contributions},
Guoquan Wang\textsuperscript{\rm 1} \footnotemark[1],
Donglin Zhou\textsuperscript{\rm 1}\thanks{Work done during an internship at Kuaishou Technology},
Sheng Yu\textsuperscript{\rm 1},
Yang Zeng\textsuperscript{\rm 1},
Wencong Zeng\textsuperscript{\rm 1},
Kun Gai\textsuperscript{\rm 2},
Guorui Zhou\textsuperscript{\rm 1} \thanks{Corresponding author}
}
\affiliations {
\textsuperscript{\rm 1}Kuaishou Technology, Beijing, China\\
\textsuperscript{\rm 2}Independent Researcher\\
\{jianghao11, wangguoquan03, zhoudonglin, yusheng03, zhengchengyi, zengwencong, zhouguorui\}@kuaishou.com, gai.kun@qq.com 
}

\usepackage{bibentry}

\begin{document}

\maketitle
\begin{abstract}
Recent advances in Large Language Models (LLMs) have enhanced text-based recommendation by enriching traditional ID-based methods with semantic generalization capabilities. Text-based methods typically encode item textual information via prompt design and generate discrete semantic IDs through item tokenization. However, in domain-specific tasks such as local-life services, simply injecting location information into prompts fails to capture fine-grained spatial characteristics and real-world distance awareness among items. To address this, we propose \textbf{LGSID}, an LLM-Aligned Geographic Item Tokenization Framework for Local-life Recommendation. This framework consists of two key components: (1) \textbf{RL-based Geographic LLM Alignment}, and (2) \textbf{Hierarchical Geographic Item Tokenization}. In the RL-based alignment module, we initially train a list-wise reward model to capture real-world spatial relationships among items. We then introduce a novel G-DPO algorithm that uses pre-trained reward model to inject generalized spatial knowledge and collaborative signals into LLMs while preserving their semantic understanding. Furthermore, we propose a hierarchical geographic item tokenization strategy, where primary tokens are derived from discrete spatial and content attributes, and residual tokens are refined using the aligned LLM's geographic representation vectors. Extensive experiments on real-world Kuaishou industry datasets show that LGSID consistently outperforms state-of-the-art discriminative and generative recommendation models. Ablation studies, visualizations, and case studies further validate its effectiveness.
\end{abstract}

\begin{links}
\link{Code}{https://github.com/JiangHaoPG11/LGSID}
\end{links}

\section{Introduction}
\begin{figure}[tp]
\includegraphics[width= \linewidth]{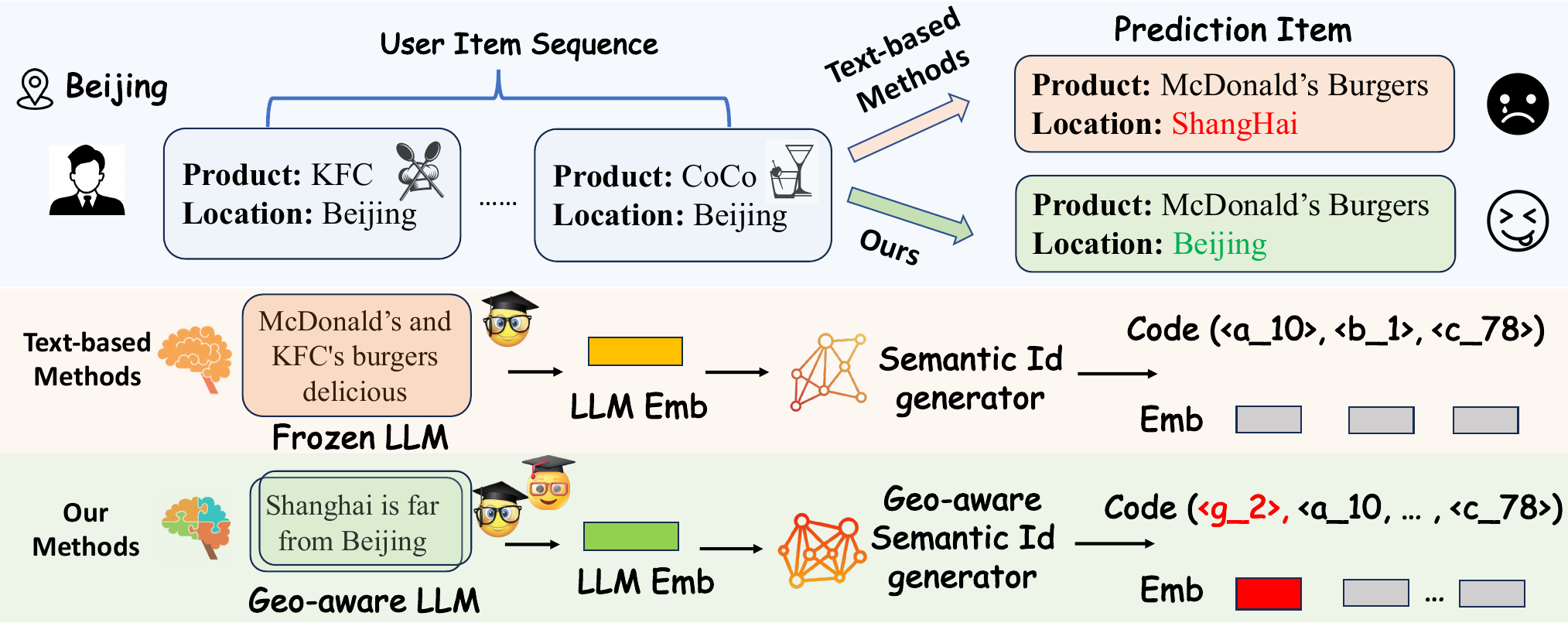}
\caption{{Illustration of challenges of text-based methods in local-life recommendation. Without geographic awareness, the system may recommend semantically relevant items that are inaccessible to users due to long distance.}}
\label{fig:f1}
\end{figure}

With the rapid growth of local-life services, recommendation systems have become essential for meeting users’ daily needs on major platforms such as Kuaishou and Meituan. However, traditional ID-based methods, which represent items using unique ID tokens and rely heavily on collaborative filtering signals \cite{he2017neural}, struggle to capture the real-world spatial characteristics and distance awareness of local-life recommendation.
In such spatially constrained scenarios \cite{ma2024successive, zhao12025spatial}, many items suffer from limited user interaction opportunities, leading to unfair exposure. Consequently, as the candidate item corpus grows, these methods encounter inherent performance bottlenecks \cite{zhang2024temporal, wu2021self}. 

Recent advances in Large Language Models (LLMs) offer an effective solution to the limited learning capacity of ID embeddings by leveraging their semantic understanding  abilities \cite{achiam2023gpt, yang2025qwen3, wang2024llmrg}.
The prevailing paradigm in text-based methods, as illustrated in the middle part of Figure~\ref{fig:f1}, involves designing prompts that incorporate domain features and item attributes for LLMs to generate semantic representations, which are subsequently quantized into discrete semantic IDs through a process known as item tokenization. For example, TIGER \cite{rajput2023recommender} is one of the first to propose the RQ-VAE quantization model, which maps LLM representations to semantic IDs (SIDs). QARM \cite{luo2024qarm} introduces the Res-Kmeans quantization model to tackle two key challenges in leveraging LLM representations: mismatch and unlearning. Furthermore, N-gram semantic IDs \cite{singh2024better, zheng2025enhancing} and Cosine semantic IDs \cite{lin2025unified} have been developed to address the long-tail problem in existing quantization methods. To better align SIDs with recommendation, several methods further inject collaborative filtering signals during or after item tokenization, while underscoring the potential of item tokenization for recommendation \cite{wang2024learnable,liu2025generative}.

Despite their successes, we argue that these methods still face two major challenges.  
(a) \textbf{Limited adaptation to downstream recommendation tasks with domain-specific knowledge.} They often treat LLMs merely as text encoders that transfer item textual descriptions into high-quality semantic representation, while focusing on designing novel quantization models to align with diverse downstream tasks. However, our key insight is that the quality and domain awareness of upstream LLMs fundamentally determine the upper bound of item tokenization performance.
(b) \textbf{Weak integration of domain knowledge and semantic understanding.} They simply combine domain-specific signals with content attributes into prompts, assuming LLMs can naturally balance their importance. However, pre-trained LLMs often misinterpret such signals and tend to prioritize content relevance over geographic proximity. As shown in Figure~\ref{fig:f1}, they may recommend a Shanghai restaurant to a user located in Beijing. This highlights the importance of item tokenization needs integrates item attributes and high-dimensional representations to better fuse semantic understanding with geographic awareness.

To overcome the limitations of existing text-based recommendation methods, we propose \textbf{LGSID}, an LLM-Aligned Geographic Item Tokenization framework for Local-life Recommendation. Unlike previous methods that focuse on quantization models, LGSID adopts a post-training strategy that aligns LLM to optimize representations, thereby equipping them with real-world spatial awareness. To this end, we integrates a novel geographic item tokenization framework through a two-stage module: (1) \textbf{RL-based Geographic LLM Alignment} and (2) \textbf{Hierarchical Geographic Item Tokenization}. The RL-based alignment module first trains reward models using distance-aware list-wise sampling to capture and compress real-world spatial relationships. Building upon this, we introduce the G-DPO algorithm to incorporate geographic and collaborative filtering signals into the LLM.  During LLM training, the G-DPO algorithm uses similarity regularization to dynamically balance semantic accuracy and geographic awareness, ensuring that domain-specific knowledge is effectively injected into the LLM while preserving its semantic understanding capabilities. Moreover, we propose a hierarchical geographic item tokenization method with a novel quantization strategy. This method first generates primary tokens based on spatial and content attributes, then refines residual tokens by leveraging the aligned LLM’s geographic representations. 
% Extensive experiments on real-world Kuaishou industry datasets demonstrate that LGSID significantly outperforms state-of-the-art discriminative and generative models. Visualization results and case study further indicate that our framework improves recall for nearby items without compromising semantic relevance, while the tokenization process enhances both geographic and semantic coherence.  

In summary, our main contributions are as follows.
\begin{itemize}[leftmargin=8pt,topsep=0pt]
\item We identify the limitations of existing LLM-driven item tokenization methods and highlight the importance of aligning LLMs with domain-specific knowledge.
\item We propose LGSID, a two-stage item tokenization framework tailored for spatially constrained scenarios such as local-life recommendation, with RL-based LLM alignment and hierarchical geographic item tokenization.
\item We conduct comprehensive experiments on real-world industry datasets, demonstrating that LGSID significantly improves performance in both discriminative and generative recommendation models.
\end{itemize}

\begin{figure*}[t]
\centering
\includegraphics[width=\textwidth]{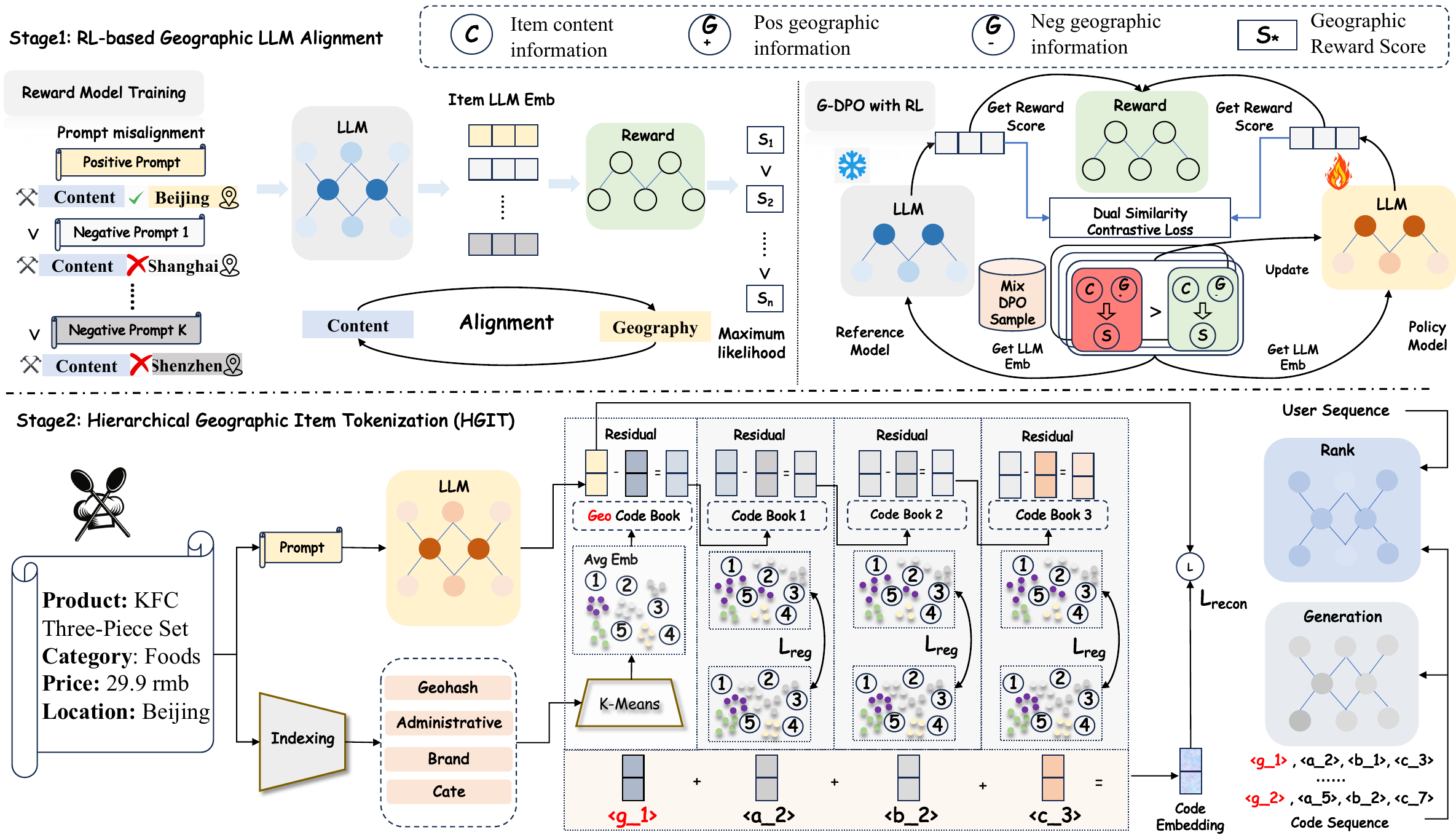}
\caption{Model structure of LGSID. The upper part illustrates the pipeline of RL-based Geographic LLM Alignment, while the lower part depicts the pipeline of Hierarchical Geographic Item Tokenization.}
\label{fig:fg2}
\end{figure*}

\section{Related Work}
\subsection{Item Tokenization}
Recent advances in item tokenization (e.g., RQ-VAE and VQ-VAE) have driven the development of text-based recommendation \cite{rajput2023recommender}. A key challenge in these methods is aligning item tokenization with recommendation. Existing item tokenization methods can be broadly categorized into two main types: two-stage methods and end-to-end methods. In the two-stage methods, early efforts such as LC-Rec~\cite{zheng2024adapting} introduces semantic alignment to integrate recommendation signals, while LETTER~\cite{wang2024learnable} jointly aligns semantic and collaborative spaces. Building on these, QARM~\cite{luo2024qarm} learns semantic IDs using Res-Kmeans guided by user interaction distributions. EAGER~\cite{wang2024eager} adopts two-stream codebooks to model both semantic and collaborative signals, UTGRec~\cite{zheng2025universal} integrates multimodal semantics with co-occurrence patterns into universal codes. Furthermore,  SC~\cite{li2025semantic} regularize semantic encoders with ID embeddings from pre-trained CF models. In end-to-end methods, tokenization and recommendation are jointly alignmented and optimized. UnifiedSID \cite{lin2025unified} employs RQ-VAE trained with cosine and euclidean distances to integrate semantic and ID tokens, while ETEGRec \cite{liu2025generative} introduces sequence-item and preference-semantic alignment objectives with generative models training. However, these methods largely overlook domain-specific constraints. For example, in local-life recommendation, users interact only with items within a limited geographic radius. SIDs without geographic awareness often recommend items that match user interests but are outside deliverable distances, negatively affects system efficiency.

\subsection{LLM Alignment for Recommendation}
A key challenge in applying LLMs to recommendation is incorporating task-specific awareness. Existing methods can be categorized into two groups. The first focuses on designing LLMs tasks to stimulate real-world knowledge and reasoning. For example, LGHRec \cite{luo2025llm} uses chain-of-thought reasoning to distill item descriptions into semantic IDs fused with vanilla IDs for GNN. GNPR-SID \cite{wang2025generative} incorporates domain attributes like location information into prompts to capture downstream signals. SIIT \cite{chen2024enhancing} iteratively refines tokenization via self-improvement. The other group focuses on fine-tuning LLMs. NoteLLM \cite{zhang2024notellm} and NoteLLM-2 \cite{zhang2024notellm2} compress items via prompts and integrate collaborative signals through supervised fine-tuning. AlignRec \cite{liu2024alignrec} introduces alignment objectives for multimodal and user–item consistency. LLMEmb \cite{liu2025llmemb} uses supervised contrastive fine-tuning to align LLM embeddings with collaborative data. LARM \cite{liu2025llm} distills open-source LLM knowledge into smaller models, while Lu et al.~\cite{lu2024aligning} strengthen LLMs’ alignment with recommendation instructions. These methods emphasize semantic alignment. However, LLMs need balance domain-specific constraints while accurately capturing users’ true preferences based on item content.

\section{Preliminaries}
\subsection{Geography Prompt Design}  
We design prompts for LLMs to derive item semantic representations by combining their textual descriptions $\mathbf{T}_i$ (e.g., name, brand, category, price) with geographical attributes (e.g., province, city, town), as detailed in Appendix: A. Specially, items are first encoded into semantic representations $\mathbf{E}_i$, which are then quantized into discrete semantic IDs $S_i$.

\subsection{SID in Recommendation}
We apply our LGSID to both two mainstream recommendation paradigms with following pipelines.
(1) {discriminative recommendation}: Users $\mathcal{U} = \{u_1, \dots, u_M\}$ and items $\mathcal{I} = \{i_1, \dots, i_N\}$ are encoded into their unique ID embeddings, where each item is represented by its textual description $\mathbf{T}_i$ and semantic IDs $S_i$. 
A user encoder $f_u(\cdot)$ learns $\mathbf{e}_u = f_u(u)$, and Top-$k$ items $\{i_1, \dots, i_k\}$ are generated by matching $\mathbf{e}_u$ with $(ID_i, \mathbf{T}_i, S_i)$. 
(2) {generative recommendation}: Items are first mapped to semantic IDs $S_i = (s_i^1, \dots, s_i^L)$ via quantized tokenization. 
Given a user's SID history $\{S_{i_1}, \dots, S_{i_t}\}$, the model predicts the next SID $S_{i_{t+1}}$, corresponding to the most likely next item.

\section{Methodology}

\subsection{Overview}
In this paper, we propose a novel geographic item tokenization framework for local-life recommendation, comprising two key modules: RL-based Geographic LLM Alignment and Hierarchical Geographic Item Tokenization. The upper part of Figure~\ref{fig:fg2} illustrates the RL-based alignment module, which aligns LLMs with real-world spatial knowledge while preserving their semantic understanding ability.
The lower part of Figure~\ref{fig:fg2} illustrates the HGIT pipeline, which fuses spatial and discrete content features with rich semantic representations derived from LLMs  to better balance item semantics and domain-specific characteristics.

\subsection{RL-based Geographic LLM Alignment}
Existing methods often incorporate domain-specific information into prompts to enhance LLMs awareness \cite{wang2025generative}. However, LLMs mainly rely on semantic similarity, capturing only coarse spatial relations from textual relevance and struggling with fine-grained distinctions between textually similar locations, e.g., confusing “Suzhou, Anhui, City” with “Suzhou, Jiangsu, City”. To address this, we propose an RL-based post-training strategy to align LLMs with real-world geographic knowledge.

\subsubsection{Geography-aware Reward Model Training.}
In real-world recommendation systems with large and dynamic candidate pools, manually labeling item pairs for reinforcement learning is impractical. To overcome this challenge, we first train a list-wise reward model \(\mathcal{R}(i)\), which predicts the geographic relevance score between item content and its corresponding location based on their LLM representations. This allows the model to internalize generalizable spatial knowledge via a neural network.

Specifically, we first compute the pairwise geodesic distance between items using their latitude and longitude, and adopt a density-aware hard negative sampling strategy that selects \(K\) negative samples based on spatial distance.
We construct prompt sequences by fixing the item content and replacing the location with that of each negative sample using a prompt mismatching strategy, as defined in Eq.~\ref{eq1}.
\begin{equation}
  \label{eq1}
  P_{i} = \bigl[P_{\text{content}}, P_{\text{location}}^{\text{i}}\bigr], \quad
  P^{j^{-}}_{i} = \bigl[P_{\text{content}}, P_{\text{location}}^{\text{j}^{-}}\bigr],
\end{equation}
where \(P_{\text{content}}\) is item text attributes, \(P_{\text{location}}^{i}\) is the true location, and \(P_{\text{location}}^{j^{-}}\) is a negative location from sampled items. This generates a prompt sequences $P_{i} = [P_{i}, P_{i}^{1}, \ldots, P_{i}^{k}]$.

The prompts are encoded by the LLM into embeddings \(\mathbf{E} = [\mathbf{E}^{i}, \mathbf{E}^{i}_{1}, \ldots, \mathbf{E}^{i}_{k}]\), where \(\mathbf{E}^{i}_{j} \in \mathbb{R}^{d}\) and \(d\) denotes the embedding dimension. 
Meanwhile, we design a list-wise architecture to facilitate the learning of spatial proximity between items. Specifically, for each prompt-mismatched sampled item, the model feeds the LLM representation $\mathbf{E}^{i}_{j}$ into a multi-layer perceptron (MLP) to predict a reward score, which quantifies the relationship between the target item content and location $j$, as defined in Eq.~\ref{eq2}.
\begin{equation}
  \label{eq2}
  r_{i,j} = \mathrm{MLP} (\mathbf{E}^{i}_{j}),
\end{equation}
where \(r_{i,j}\) denotes the reward score between the $i$-th target item and $j$-th sampled item.
Next, we assign soft labels \( p_{i,j} \) based on the distances between items. Specifically, given a candidate list of \( K \) prompt sequence sorted by distance from near to far, we define the soft labels as shown in Eq.~\ref{eq333}.
\begin{equation}
\label{eq333}
p_{i,j} = K - Rank(\text{dis}_{i,j}) + 1,
\end{equation}
where \( \text{dis}_{i,j}\) denotes the haversine distance between items \(i\) and \(j\) in the list. This distance-based labeling strategy encourages the model to prioritize items that are geographically closer to the target. The reward model is trained using a weighted binary cross-entropy loss, as defined in Eq.~\ref{eq3}. 
\begin{equation}
  \label{eq3}
  \mathcal{L}_{\mathrm{RM}} = - \frac{1}{N} \sum_{i=1}^N \sum_{j} p_{i,j} \log \sigma (r_{i,j}),
\end{equation}
where \(\sigma(\cdot)\) is the sigmoid activation function, \(N\) is the batch size, and \(\mathcal{L}_{\mathrm{RM}}\) denotes the reward model’s training loss.

\subsubsection{G-DPO Algorithm with RM.}
Building upon pre-trained reward models, we propose G-DPO, an enhanced algorithm inspired by Direct Preference Optimization (DPO) \cite{rafailov2023direct}, as illustrated in the upper right part of Figure~\ref{fig:fg2}. In G-DPO, we introduce a domain-mixed sampling strategy for preference learning, denoted by \(\mathcal{D}_{\text{mix}} = \mathcal{D}_{\text{dc}} \cup \mathcal{D}_{\text{gc}}\). This database combines two data types: domain collaborative pairs and geography constrained pairs. 

For domain collaborative pairs $\mathcal{D}_{\text{dc}}$, we utilize user historical behaviors to enhance collaborative signal awareness. Items frequently co-occurrence by users tend to exhibit both semantic and geographic similarity \cite{zhang2024notellm}. Specifically, the co-occurrence score is defined in Eq.~\ref{eq4}.
\begin{equation}
\label{eq4}
  s_{i_a, i_b} = \sum_{u=1}^{U} \mathbb{I} \bigl[ i_a \in H_u \wedge i_b \in H_u \bigr],
\end{equation}
where \(\mathbb{I}[\cdot]\) is the indicator function, and \(H_u\) represents user's click history. We retain pairs \((i_a, i_b)\) with \(s_{i_a, i_b} > s_{\mathrm{th}}\) as domain collaborative sample pairs, where \(s_{\mathrm{th}}\) is the threshold.

For geographically constrained pairs $\mathcal{D}_{\text{gc}}$, we randomly sample items outside the target item's location to form pairs \((i_a, i_r)\), ensuring diversity and efficiency within million-scale candidate pools. Furthermore, G-DPO aligns the policy model \(\pi_\theta\) with a reference model \(\pi_{\text{ref}}\) using domain-mixed sample pairs \((i^+, i^-) \in \mathcal{D}_{\text{mix}}\). Each item pair is scored by the reward model \(\mathcal{R}\), which takes embeddings from either the policy \(\mathbf{E}_{\pi_\theta}(i)\) or the reference model \(\mathbf{E}_{\text{ref}}(i)\) and outputs a distance-based score. The alignment loss in G-DPO is given in Eq.~\ref{eq5}.
\begin{align}
\label{eq5}
\mathcal{L}_{\text{align}} = 
- \mathbb{E}_{(i^+, i^-)} \log \sigma \Big( \beta \Big(
\mathcal{R}( \mathbf{E}_{\pi_\theta}(i^+)) - \mathcal{R}(E_{\pi_\theta}(i^-)) \notag \\
- \mathcal{R}(\mathbf{E}_{\pi_{\text{ref}}}(i^+)) 
+ \mathcal{R}(\mathbf{E}_{\pi_{\text{ref}}}(i^-))
\Big) \Big),
\end{align}
where $\beta$ controls alignment sharpness, $\sigma(\cdot)$ is the sigmoid function, and \(\mathcal{R}(\cdot)\) is the pre-trained reward model. 
To preserve LLMs' semantic representations during G-DPO updates, we introduce an {in-batch contrastive loss} as a similarity regularizer, as shown in Eq.~\ref{eq6}. 
\begin{equation}
\label{eq6}
\begin{aligned}
  \mathcal{L}_{\text{sim}} = \mathbb{E}_{i \in \mathcal{B}} \Big[ 
  &\| E_{\pi_{\theta}}(i) - E_{\pi_{\text{ref}}}(i) \|_2^2 \\
  &- \frac{1}{|\mathcal{B}| - 1} \sum_{j \ne i} \| E_{\pi_{\theta}}(i) - E_{\pi_{\text{ref}}}(j) \|_2^2 
  \Big].
\end{aligned}
\end{equation}
For each instance \(i\) in a batch \(\mathcal{B}\), we pull its representation closer to the reference model output while pushing it away from other reference embeddings in the same batch.
We combine the contrastive and DPO loss with weight \(\lambda\), as shown in Eq.~\ref{eq7}.
\begin{equation}
  \label{eq7}
  \mathcal{L}_{\text{G-DPO}} = \mathcal{L}_{\text{align}} + \lambda \mathcal{L}_{\text{sim}}.
\end{equation}

\subsection{Hierarchical Geographic Item Tokenization}

We introduce a Hierarchical Geographic Item Tokenization module on top of aligned LLM semantic representations.

For the first layer, we construct a multi-dimensional feature that integrates different types of attributes for geography-aware token initialization. 
To mitigate the inefficiency of high-dimensional one-hot encoding, we adopt composite embeddings for discrete categorical features. Specifically, geography-aware codes $f_{\text{geo}}$ are formed by latitude and longitude. Meanwhile, administrative codes $f_{\text{admin}}$ are fixed scaling factors used for normalization, determined by the province ID, city ID, and district ID. Similarly, category codes $f_{\text{cat}}$ are fixed scaling factors based on the primary category and the secondary category. Brand codes $f_{\text{brand}}$ are fixed scaling factors determined by the brand ID. The final clustering feature vector \(\mathbf{F}\) is constructed via weighted concatenation of all components, as shown in Eq.~\ref{eq10}.
\begin{equation}
\label{eq10}
\mathbf{F} = [w_{\text{admin}} \cdot f_{\text{admin}},\; w_{\text{geo}} \cdot f_{\text{geo}},\; w_{\text{cat}} \cdot f_{\text{cat}},\; w_{\text{brand}} \cdot f_{\text{brand}}],
\end{equation}
where \(w_*\) are empirically chosen coefficients reflecting the relative importance of each feature type. Using clustering feature vector \(\mathbf{F}\), we apply MiniBatch K-Means to generate a vocabulary of first-layer geographic tokens. The cluster representations are then computed as the mean of LLM embeddings within each token group, producing the first-layer cluster centers \(\boldsymbol{\mu}^{(1)}\).

For residual layers ($l \geq 2$), we employ learnable cluster centers with euclidean distance-based assignment. The process of the cluster centers $\boldsymbol{\mu}^{(l)}$ with Eq.~\ref{eqrun}, while the quantized representation and residual update with Eq.~\ref{eq11} and \ref{eq11-2}.
\begin{equation}
\label{eqrun}
\mathbf{z}^{(l)} = \arg\min_k \|\mathbf{R}^{(l-1)} - \boldsymbol{\mu}^{(l)}\|_2^2,
\end{equation}
\begin{equation}
\label{eq11}
\mathbf{Q}^{(l)} = \boldsymbol{\mu}^{(l)}[\mathbf{z}^{(l)}], 
\end{equation}
\begin{equation}
\label{eq11-2}
\mathbf{R}^{(l)} = \mathbf{R}^{(l-1)} - \mathbf{Q}^{(l)}.
\end{equation}

The primary objective is to minimize the reconstruction loss between the original embeddings and their quantized representations. For input embeddings $\mathbf{X}$, we apply a combination of absolute reconstruction losses with Eq.~\ref{eq13}.
\begin{equation}
\label{eq13}
\mathcal{L}_{\text{recon}} = \|\mathbf{{X}} - \sum_{l=1}^L \mathbf{{Q}}^{(l)}\|_2^2. 
\end{equation}

To promote balanced utilization of the learned clusters and prevent cluster collapse, we introduce an entropy-based regularization term. For each layer $l$, the cluster usage distribution is computed as shown in Eq.~\ref{eq14}.
\begin{equation}
\label{eq14}
p_k^{(l)} = \frac{1}{N} \sum_{i=1}^N \mathbb{I}[\mathbf{z}_i^{(l)} = k],
\end{equation}
where $\mathbb{I}[\cdot]$ is the indicator function. The regularization loss encourages uniform cluster usage through KL divergence:
\begin{equation}
\mathcal{L}_{\text{reg}}^{(l)} = \text{KL}\left(p^{(l)} \| \mathbf{u}\right) = \sum_{k=1}^{K_l} p_k^{(l)} \log \frac{p_k^{(l)}}{1/K_l},
\end{equation}
where $\mathbf{u} = [1/K_l, \ldots, 1/K_l]$ represents the uniform distribution over $K_l$ clusters. The complete training objective combines reconstruction with cluster usage regularization:
\begin{equation}
\mathcal{L}_{\text{HGIT}} = \mathcal{L}_{\text{recon}} + \lambda_{\text{reg}} \sum_{l=2}^L \mathcal{L}_{\text{reg}}^{(l)},
\end{equation}
where $\lambda_{\text{reg}}$ controls the strength of the regularization. Note that regularization is only applied to learnable layers ($l \geq 2$) since the first layer uses pre-computed geographic clusters.

\section{Experiment}
In this section, we conduct extensive experiments on real-world datasets to address the following research questions:
\begin{itemize}[leftmargin=10pt,topsep=0pt]
\item \textbf{RQ1}: How does LGSID improve upon existing tokenization for SOTA discriminative and generative models?
\item \textbf{RQ2}: How does fine-tuning affect the LLM’s ability on enhancing geographic awareness while preserving semantic understanding?
\item \textbf{RQ3}: How accurately do item representations and quantized IDs reflect real-world geographic proximity?
\item \textbf{RQ4}: How does LGSID exhibit superior geographic awareness compared to other SID methods?
\end{itemize}

\subsection{Experiment Settings}
The datsets and evaluation, finetune settings, parameters settings and baseline models are detailed in Appendix: B.

\subsection{Overall Performance (RQ1)}

\begin{table*}[t]
\centering
\footnotesize
\begin{tabular}{l|ccccc}
\toprule
\textbf{Method} & \textbf{DIN} & \textbf{DIEN} & \textbf{SIM} & \textbf{TWIN} & \textbf{ETA} \\
\midrule
Origin & 0.5859 & 0.6255 & 0.5884 & 0.5898 & 0.5903 \\
+ Res-KMeans \cite{luo2024qarm} & 0.6100\tiny{$\uparrow$+0.0241} & 0.6369\tiny{$\uparrow$+0.0114} & 0.6063\tiny{$\uparrow$+0.0179} & 0.6087\tiny{$\uparrow$+0.0189} & 0.6077\tiny{$\uparrow$+0.0174} \\
+ RQ-VAE \cite{rajput2023recommender} & 0.6185\tiny{$\uparrow$+0.0326} & 0.6364\tiny{$\uparrow$+0.0109} & 0.6111\tiny{$\uparrow$+0.0227} & 0.6153\tiny{$\uparrow$+0.0255} & 0.6153\tiny{$\uparrow$+0.0250} \\
+ Lin et al. \cite{lin2025unified} & 0.6161\tiny{$\uparrow$+0.0302} & 0.6368\tiny{$\uparrow$+0.0113} & 0.6107\tiny{$\uparrow$+0.0223} & 0.6148\tiny{$\uparrow$+0.0250} & 0.6148\tiny{$\uparrow$+0.0245} \\
+ RQ-VAE-ngram \cite{zheng2025enhancing} & 0.6163\tiny{$\uparrow$+0.0304} & 0.6354\tiny{$\uparrow$+0.0099} & 0.6116\tiny{$\uparrow$+0.0232} & 0.6129\tiny{$\uparrow$+0.0231} & 0.6145\tiny{$\uparrow$+0.0242} \\
+ LGSID (Ours) & \textbf{0.6276\tiny{$\uparrow$+0.0417}} & \textbf{0.6484\tiny{$\uparrow$+0.0229}} & \textbf{0.6224\tiny{$\uparrow$+0.0340}} & \textbf{0.6263\tiny{$\uparrow$+0.0365}} & \textbf{0.6274\tiny{$\uparrow$+0.0371}} \\
\bottomrule
\end{tabular}
\caption{Performance comparison with AUC improvements over the original discriminative recommendation models.}
\label{tab:auc1}
\end{table*}

\begin{table*}[t]
\centering
\footnotesize
\resizebox{\textwidth}{!}{ % 自动缩放至文本宽度
\begin{tabular}{l|cccc|cccc}
\toprule
\textbf{Method} & \multicolumn{4}{c|}{\textbf{TIGER}} & \multicolumn{4}{c}{\textbf{OneRec}} \\
\cmidrule(lr){2-5} \cmidrule(lr){6-9}
& Hit@5 & Hit@10 & NDCG@5 & NDCG@10 & Hit@5 & Hit@10 & NDCG@5 & NDCG@10 \\
\midrule
RQ-VAE \cite{rajput2023recommender} & 0.3087 & 0.3880 & 0.2255  & 0.2512 & 0.3739 &  0.4534 & 0.2798 & 0.3056 \\
Lin et al. \cite{lin2025unified} & 0.1767 & 0.2067 & 0.1335 & 0.1432 & 0.2950 & 0.3346 & 0.2272 & 0.2401 \\
RQ-VAE-ngram \cite{zheng2025enhancing} & 0.2991 & 0.3769 & 0.2158 & 0.2411 & 0.3626 & 0.4358 & 0.2720 & 0.2957 \\
LGSID (Ours) & \textbf{0.3921} & \textbf{0.5077} & \textbf{0.2817} & \textbf{0.3191} & \textbf{0.4435} & \textbf{0.5537} & \textbf{0.3304} & \textbf{0.3661} \\
\midrule
\textbf{IMP} & 
\textbf{27.01}\% & 
\textbf{30.83}\% & 
\textbf{24.94}\% & 
\textbf{27.05}\% & 
\textbf{18.63}\% & 
\textbf{22.13}\% & 
\textbf{18.09}\% & 
\textbf{19.79}\% \\
\bottomrule
\end{tabular}
}
\caption{Performance comparison with different quantization methods in generative recommendation models. The \textbf{IMP} metric indicates the relative improvement of LGSID over the best-performing baseline (excluding LGSID).}
\label{tab:gs}
\end{table*}

\subsubsection{The results in Discriminative Recommendation.}
Table \ref{tab:auc1} reports offline AUC on the Kuaishou local-life dataset when DIN \cite{zhou2018deep}, DIEN \cite{zhou2019deep}, ETA \cite{chen2021end}, SIM \cite{pi2020search}, and TWIN \cite{si2024twin} are augmented with different item tokenization schemes.  LGSID delivers the largest absolute gains.
These improvements stem from the two components of LGSID. For attention-based DIN, DIEN and SIM, flat item IDs severely limit the granularity at which spatial proximity can influence attention scores.  After the G-DPO phase of LGSID injects aligned LLM's spatial knowledge, each ID becomes a geography-aware embedding that encodes real-world spatial distance and neighborhood co-visit patterns.  
%Consequently, DIN’s attention layer reallocates weight toward truly nearby items, driving its AUC 4.17\% above the original baseline.  DIEN and SIM exhibit similar behaviour, rising 2.29\% and 3.40\% respectively.
For TWIN and ETA, which operate under tight latency constraints, compact yet informative codes are critical. LGSID’s hierarchical quantization first compresses geo-textual attributes into coarse primary tokens, then progressively refines residuals with geographic context.  This yields richer representations without expanding the embedding table, lifting TWIN by 3.65\% and ETA by 3.71\%.  

\subsubsection{The results in Generative Recommendation.}
Table \ref{tab:gs} shows the results of generative recommendation. We mainly compare two generative recommendation model, including TIGER ~\cite{rajput2023recommender} and OneRec \cite{deng2025onerec} with different quantization methods. We find that Lin et al. \cite{lin2025unified} obtains the worst performance, possibly because it uses different distance functions across various levels of codewords, leading to convergence difficulties and challenges in model optimization. RQ-VAE \cite{rajput2023recommender} and RQ-VAE-ngram \cite{zheng2025enhancing} achieve similar performance on TIGER and OneRec. However, these quantization methods did not consider geographical constraints, resulting in suboptimal performance in local-life recommendation scenarios. Our LGSID introduces RL-based Alignment to generate geographically aware representations and transfers these representations into semantic IDs through hierarchical geographic item tokenization.

\subsection{RL-based LLM Alignment Analysis (RQ2)}
\begin{table*}[tp]
\centering
\resizebox{\textwidth}{!}{
\begin{tabular}{l|ccc|ccc|ccc|ccc}
\toprule
\textbf{Method}
& \multicolumn{3}{c|}{\textbf{Similarity}} 
& \multicolumn{3}{c|}{\textbf{Province Coverage (P@K)}} 
& \multicolumn{3}{c|}{\textbf{City Coverage (C@K)}} 
& \multicolumn{3}{c}{\textbf{Town Coverage (T@K)}} \\
\cmidrule(lr){2-4} \cmidrule(lr){5-7} \cmidrule(lr){8-10} \cmidrule(lr){11-13}
 & Top@5 & Top@10 & Top@100 & P@5 & P@10 & P@100 & C@5 & C@10 & C@100 & T@5 & T@10 & T@100 \\
\midrule
Origin & 0.9204 & 0.9133 & 0.8833 & 0.8716 & 0.8410 & 0.6681 & 0.7342 & 0.6827 & 0.4372 & 0.1601 & 0.1328 & 0.0552 \\
DPO-PR & 0.8771 & 0.8679 & 0.8286 & 0.9001 & 0.8752 & 0.7477 & 0.7478 & 0.7013 & 0.5064 & 0.1452 & 0.1167 & 0.0445 \\
DPO-LR &  0.7595 & 0.7478	& 0.7088  & 0.8995 & 0.8648 & 0.6560 & 0.8681 & 0.8254 & 0.5783 & 0.5584  & 0.4966 & 0.2480 \\
DPO-LRD & 0.7411 & 0.7288 & 0.6876 & 0.8715 & 0.8302 & 0.6012 & 0.8277 & 0.7755 & 0.5043 & 0.6114 & 0.5435 & 0.2620 \\
DPO-LRDM & 0.8107 & 0.7954 & 0.7401 & 0.9047 & 0.8773 & 0.7261 & 0.7812 & 0.7329 & 0.5218 & 0.1816 & 0.1481 & 0.0625 \\
DPO-LRDMS & 0.8856 & 0.8754 & 0.8283 & 0.9960 & 0.9936 & 0.9662 & 0.9548 & 0.9352 & 0.8130 & 0.4030 & 0.3525 & 0.2260 \\
G-DPO (Ours) & 0.8977 & 0.8892 & 0.8504 & 0.9905 & 0.9852 & 0.9307 & 0.9173 & 0.8858 & 0.7065 & 0.294 & 0.2432 & 0.1290 \\
\midrule
\textbf{IMP} & \textbf{-2.47\%} & \textbf{-2.64\%} & \textbf{-3.72\%} & \textbf{+13.64\%} & \textbf{+17.15\%} & \textbf{+39.31\%} & \textbf{+24.94\%} & \textbf{+29.75\%} & \textbf{+61.60\%} & \textbf{+83.64\%} & \textbf{+83.13\%} & \textbf{+133.70\%} \\
\bottomrule
\end{tabular}
}
\caption{Evaluation of G-DPO variants: Origin (baseline), DPO-PR (point-wise reward), DPO-LR (list-wise reward), DPO-LRD (list-wise reward + density-aware sampling), DPO-LRDM (list-wise reward + density-aware sampling + domain mixed preference pairs), and DPO-LRDMS (list-wise reward + density-aware sampling + domain mixed preference pairs + similarity regularization). Metrics include similarity, province (P@K), city (C@K), and town (T@K) coverage for \(K = \{5, 10, 100\}\).}
\label{tab:gdpo_variants}
\end{table*}

This experiment validates the effectiveness of our RL-based LLM Alignment by measuring model improvements before and after fine-tuning. We design two metrics based on retrieving the top-\(k\) items by similarity to a target item embedding: (1) \textbf{semantic similarity}, measured by the semantic similarity of the retrieved items, and (2) \textbf{geographic awareness}, measured by the coverage of retrieved items sharing the same province (state), city, and town as the target item. 
Based on Table~\ref{tab:gdpo_variants}, we have following conclusion. (1) Pure semantic understanding is insufficient for geographic awareness, as text similarity captures terms but not real distances. (2) The reward model effectively compresses and transfers geographic knowledge into the LLM, where list-wise modeling improves T@5 from 0.1601 to 0.5584. (3) Incorporating density-aware list-wise modeling further boosts T@5  from 0.1601 to 0.6114, enhancing near-distance sensitivity. However, this comes at the expense of semantic comprehension; applying a mixed-sample strategy mitigates this trade-off by improving sample discrimination and integrating collaborative signals. (4) Over emphasizing geographic awareness alone does not ensure better downstream performance, so we introduce textual similarity regularization to maintain semantics while achieving optimal results.

\begin{figure}[t]
\includegraphics[width= \linewidth]{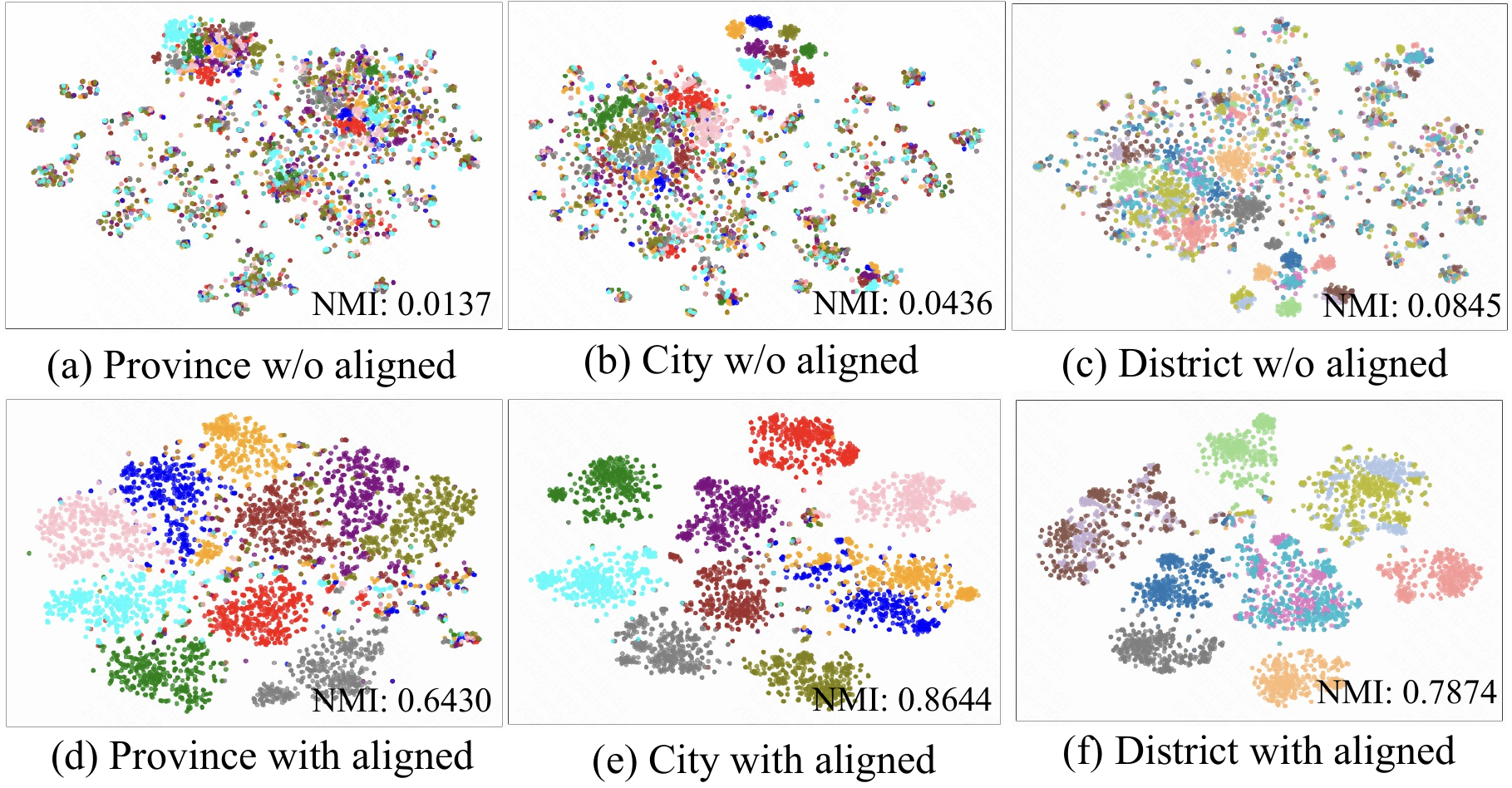}
\caption{{T-SNE visualization of items around cluster centroids across tokenization methods.}}
\label{f3p}
\end{figure}

\begin{figure}[t]
\includegraphics[width= \linewidth]{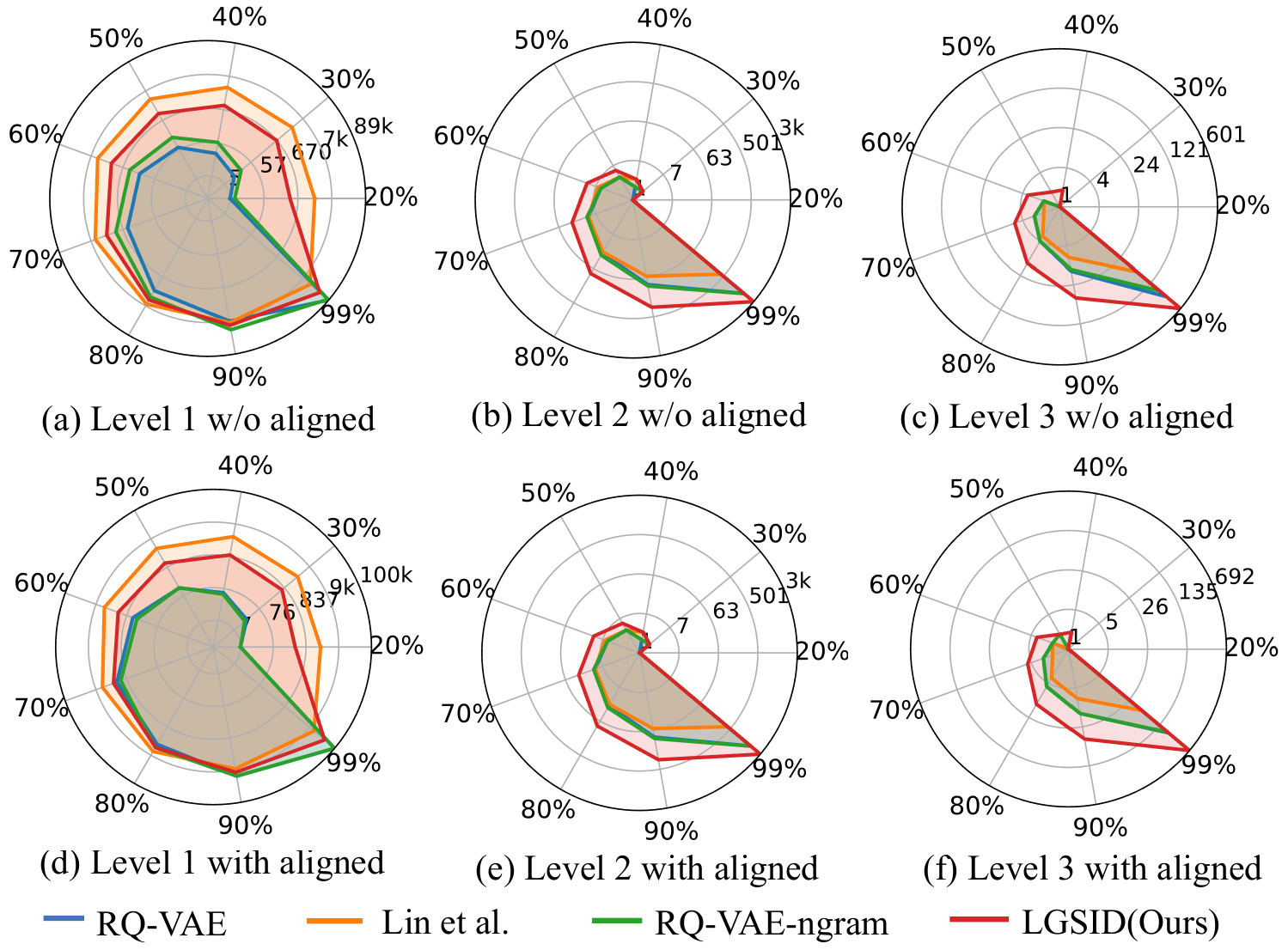}
\caption{{Token quantile percentiles across hierarchical levels for local-life items.}}
\label{f2p}
\end{figure}
\begin{figure*}[ht]
\includegraphics[width= \linewidth]{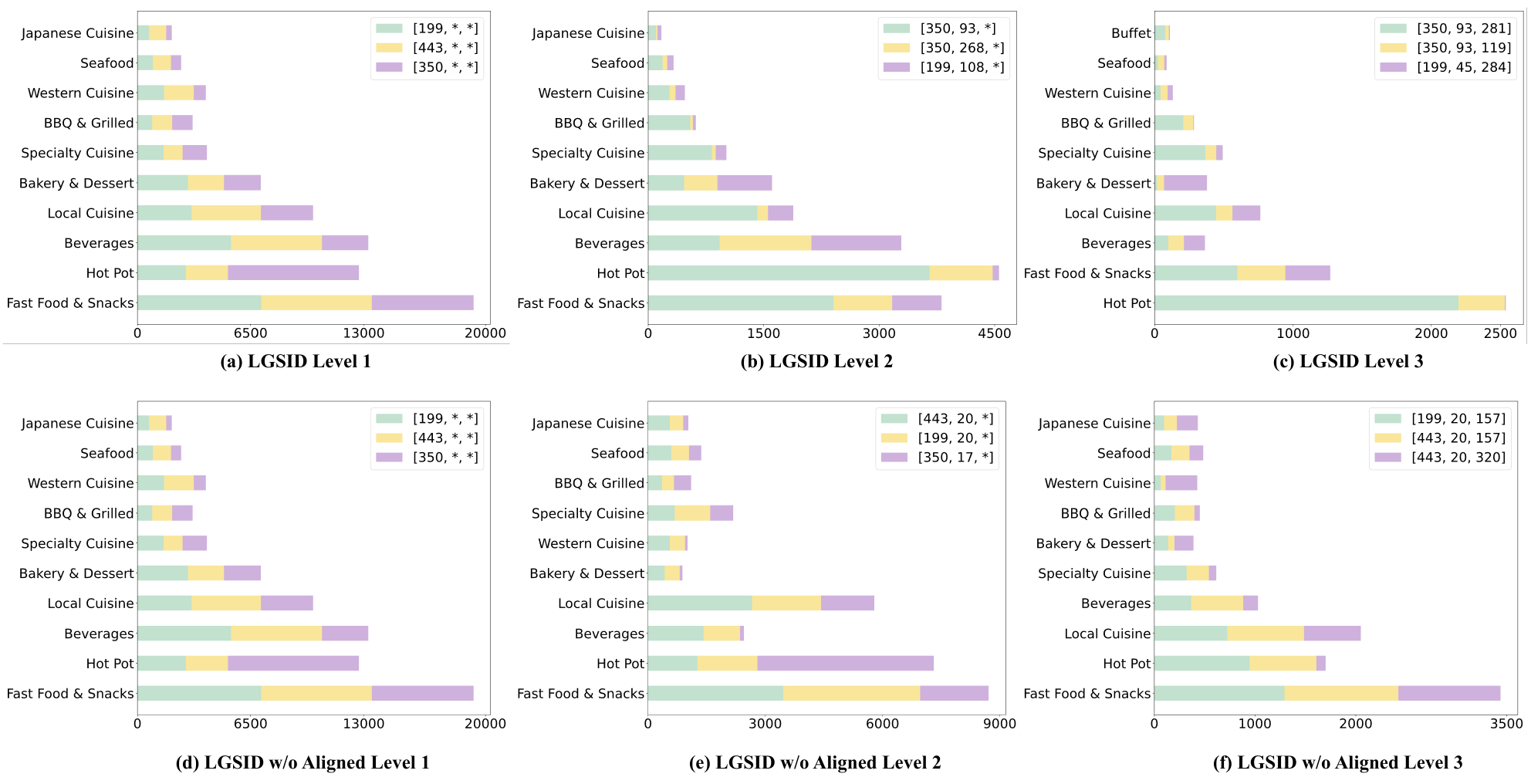}
\caption{{Hierarchical category frequency distribution of LGSID for different SID prefixes (Aligned vs Unaligned).}}
\label{f0p}
\end{figure*}

\subsection{Visualization Analysis (RQ3)}
The T-SNE visualization in Figure \ref{f3p} shows that after RL-based Geographic LLM Alignment, the cluster centers of Province, City, and District are significantly closer.
The Normalized Mutual Information (NMI) quantitatively measures the agreement between the clustering partition discovered by the model and the ground-truth geographic labels.
It jumps from 0.0137–0.0845 to 0.6430–0.8644.
The key to this improvement lies in: G-DPO first uses distance-aware list-wise rewards to inject real-world spatial relationships into the LLM, allowing its token embeddings to carry inherent geographic priors. 

%Meanwhile, the proposed method shifts the cross-domain adaptation burden to the LLM, strengthening geographic awareness while preserving semantic discriminability, providing a tighter performance upper bound for downstream recommendation tasks.

The radar chart in Figure \ref{f2p} shows the quantile values at different percentile levels, where larger areas indicate superior coverage performance across the distribution spectrum. Higher quantile values indicate that tokens can represent more instances. 
At Level-1, our LGSID method demonstrates remarkable consistency between aligned and non-aligned settings, with identical coverage patterns observed in both scenarios.  LGSID maintains 11k coverage in 90\% quantile, while RQ-VAE decay to 8k. 
As we progress to finer granularities (Level-2 and Level-3), LGSID's advantages become more pronounced. Its area in the radar chart is the largest compared to other methods. 

\subsection{Case Study (RQ4)}
Figures \ref{f0p} show the allocation of three-layer discrete tokens of the LGSID hierarchical quantizer with and without RL-based G-DPO alignment, respectively. 
Since the first layer uses pre-computed geographic clusters, the distribution of categories in Figure \ref{f0p}(a) alignment and Figure \ref{f0p}(d) misalignment is similar.
Figure \ref{f0p}(b) shows that after RL-based alignment the first-layer token [350, 93, *] cleanly groups the entire BBQ \& Grilled branch into one coarse identifier.  In contrast, Figure \ref{f0p}(e) presents the same layer without alignment: the same restaurants are now scattered across [199, 20, *], [443, 20, *] and [350, 17, *], because the LLM embeddings have not been post-trained with G-DPO to respect distance-aware rewards.  As a result, the un-aligned Level-1 tokens lose category cohesion; the hierarchical quantizer can no longer rely on a shared root to refine sub-categories in later levels, illustrating the importance of upstream LLMs and their quality of embeddings.

\section{Conclusion}

In this paper, LGSID is designed to equip semantic IDs with real-world spatial awareness for local-life recommendation. Specifically, LGSID trains a list-wise reward model with density-aware negative sampling to capture relative spatial distances, then injects geographic knowledge using a novel G-DPO algorithm. Moreover, a hierarchical geographic tokenization strategy generates a sequence of spatial-aware discrete tokens, enabling efficient compression and reconstruction. Extensive results on both discriminative and generative recommendation models demonstrate improved geographic awareness in scenarios of different granularity. 

% \section{Acknowledgments}
\bibliography{ref}

\clearpage
\appendix
\section{Appendix A: Prompts Design in Local-life Recommendation}

\begin{figure}[h]
\includegraphics[width= \linewidth]{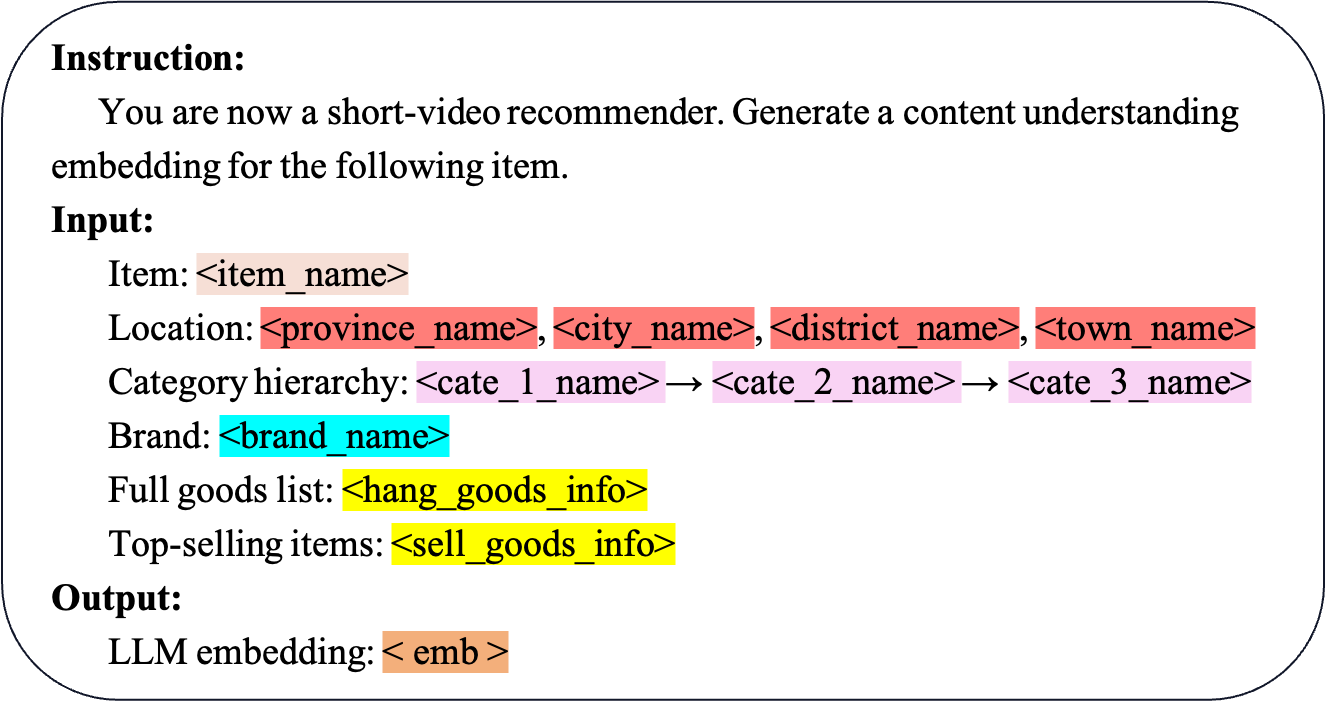}
\caption{{Illustration of our prompt design.}}
\label{f3}
\label{prompts}
\end{figure}

To obtain the LLM representation of each item in local-life recommendation, we design a prompt that incorporates domain-specific features and item content, as illustrated in the Figure \ref{prompts}.

\label{appendix：A}

\section{Appendix B: Experiment Settings}

\subsubsection{Datasets and evaluation metrics}
We conduct extensive experiments on real-world industry dataset from the feed of the Kuaishou App in the local-life scenario, which are summarized in Table \ref{data}. It contains all items include both item geographic and textual information
Public datasets that satisfy delivery-distance constraints and contain detailed item text and geographic location information are scarce. Therefore, we only use an industrial dataset in this study. In future work, we plan to release our dataset to support research on algorithms for LBS-like recommendation scenarios. 
We adopt task-specific preprocessing. For discriminative models, all positives are retained while negatives are down-sampled at a 1:4 ratio, with AUC~\cite{hanley1982meaning} as the metric. For generative models, we maintain the down-sampled results and follow a leave-one-out split~\cite{rajput2023recommender}: the last item for testing, second-last for validation, and the rest for training. Evaluation uses NDCG@K and Hit@K with \(k = \{5, 10\}\)~\cite{jarvelin2002cumulated}.

% \footnote{we will release our dataset upon acceptance to support further research.}

\begin{table}[h]
\centering
\begin{tabular}{lccc}
\toprule
\multicolumn{4}{c}{\textbf{Kuaishou Industry Datasets}} \\
\midrule
\# Samples    & 50,000,000 & \# Users      & 19,080,888 \\
\# Items      & 2,325,266  & \# Brands     & 19,408 \\
\# Categories & 818         \\
\bottomrule
\end{tabular}
\caption{Statistics of the Kuaishou local-life dataset.}
\label{data}
\end{table}

\subsection{Finetune Settings}
We employ \textbf{BGE} \cite{bge_embedding} as our backbone, a state-of-the-art multilingual text understanding model. BGE is well-suited for industrial recommendation scenarios as it requires no complex instructions. Following the standard protocol, we set the prompt length to 512 tokens and the embedding size to 1024, using the last token’s hidden state to represent the entire text. For the {G-DPO} algorithm, each positive sample is paired with 15 negative samples to construct list-wise inputs for the reward model, which consists of a two-layer MLP with a sigmoid activation. For effective LLM fine-tuning, we adopt the {LoRA} \cite{hu2022lora} with a rank of 8 and a dropout rate of 0.05, fine-tuning only the key and value layers to preserve semantic understanding. In the {DPO loss}, we set $\beta = 0.9$, use Euclidean distance for the similarity constraint with a weight of $\lambda = 155$, and adopt in-batch contrastive learning. For {domain-mixed DPO preference pairs}, we filter samples using a co-occurrence score threshold of 1200 and randomly select one negative item per sample to construct the training pairs.

\subsection{Parameters Settings}
In our experimental setup, we set batch size to 10,240, and the optimal settings are selected based on model performance on the validation data. We set the embedding dimension for each feature to 8, with the predict MLP tower dimensions configured as [32, 16, 1]. We employ AdamW as the optimizer, applying learning rates of 0.1, with a steplr scheduler that decays the learning rate by a factor of 0.9 every 500 steps. All experiments are conducted on two GPUs, each equipped with 48GB of memory.
In practice, we first train the reward model, then perform LLM alignment using the G-DPO algorithm, and finally train the quantization model.

\subsection{Baselines}
To evaluate the effectiveness of our model, we compare it with the following representative baselines.  
For {discriminative recommendation}, we select DIN \cite{zhou2018deep}, DIEN \cite{zhou2019deep}, ETA \cite{chen2021end}, SIM \cite{pi2020search}, and TWIN \cite{si2024twin}, which focus on learning effective user-item interaction representations and are widely adopted in industrial applications. For {generative recommendation}, we choose these two as representative model such as Tiger \cite{rajput2023recommender} and OneRec \cite{deng2025onerec}.

\section{Appendix C: Visualization Analysis}
\begin{figure*}[tp]
\includegraphics[width= \linewidth]{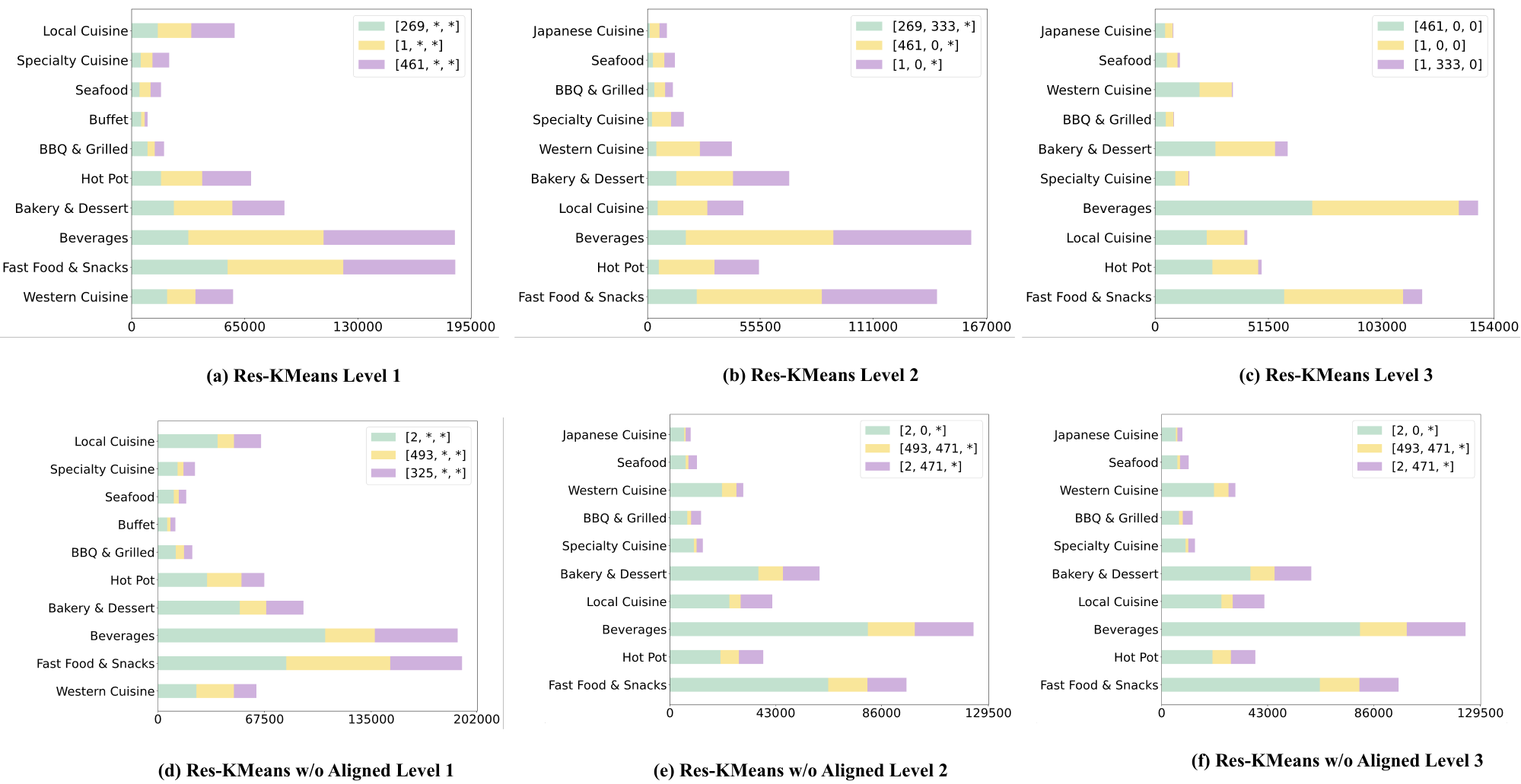}
\caption{{Hierarchical category frequency distribution of Res-KMeans for different SID prefixes (Aligned vs Unaligned).}}
\label{f0}
\end{figure*}

Figure \ref{f0} shows the Level-1, Level-2, and Level-3 token assignments produced by the Res-KMeans hierarchical quantizer both with and without G-DPO alignment.  Pre-computed geographic clusters still govern this layer, so the overall cuisine distribution—Local Cuisine, Specialty Cuisine, Seafood, Buffet, BBQ \& Grilled, Hot Pot, Bakery \& Dessert, Beverages, Fast Food \& Snacks, Western Cuisine—remains visually similar to its un-aligned counterpart.  Critically, the aligned token [269, *, *] unifies the entire Local Cuisine branch, while [461, *, *] cleanly gathers Specialty Cuisine, giving Level-2 a single, coherent root from which to refine sub-categories and illustrating that distance-aware rewards preserve category cohesion.

\begin{figure*}[tp]
\includegraphics[width= \linewidth]{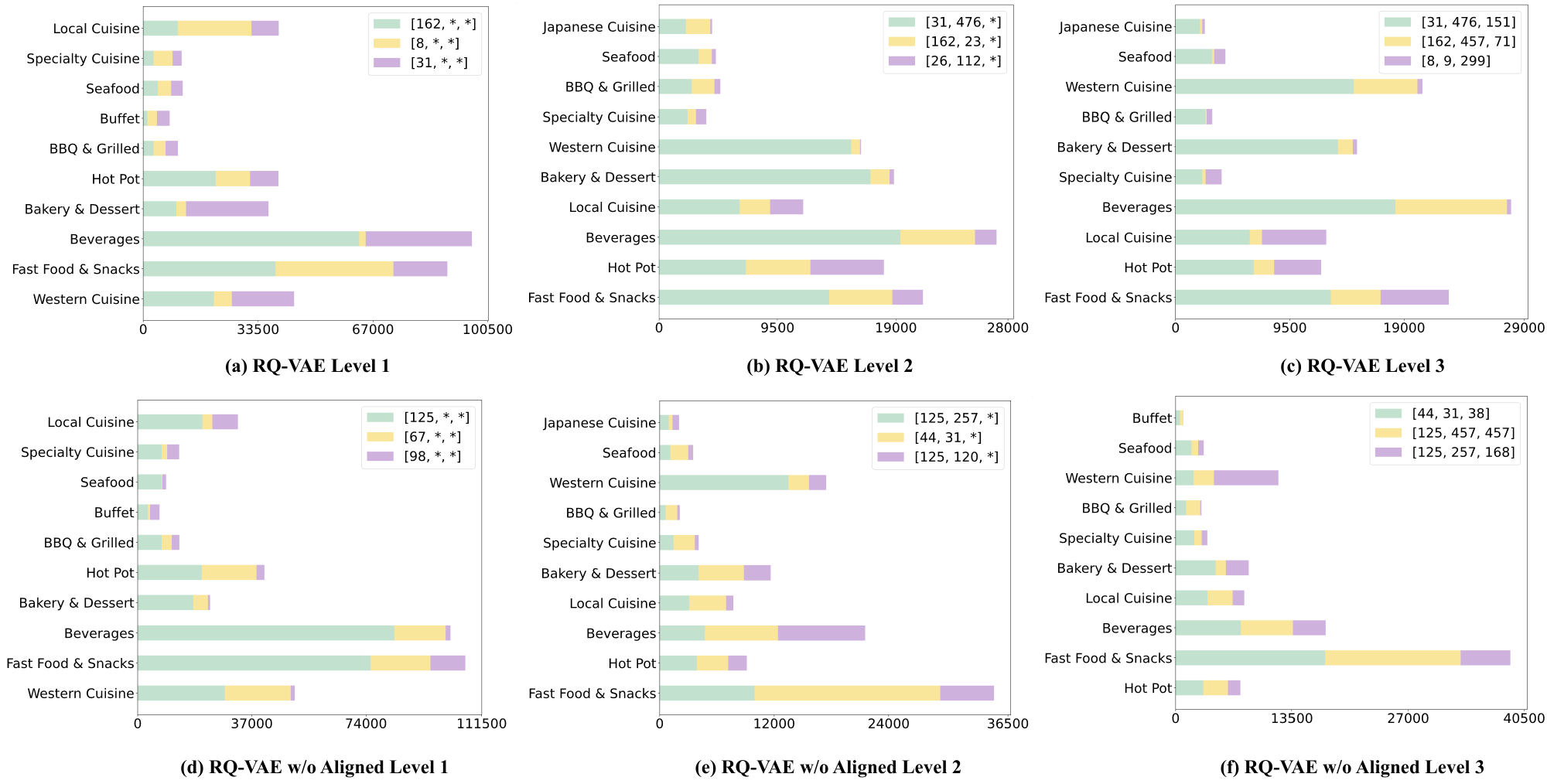}
\caption{{Hierarchical category frequency distribution of RQ-VAE for different SID prefixes (Aligned vs Unaligned).}}
\label{f19}
\end{figure*}

\begin{figure*}[tp]
\includegraphics[width= \linewidth]{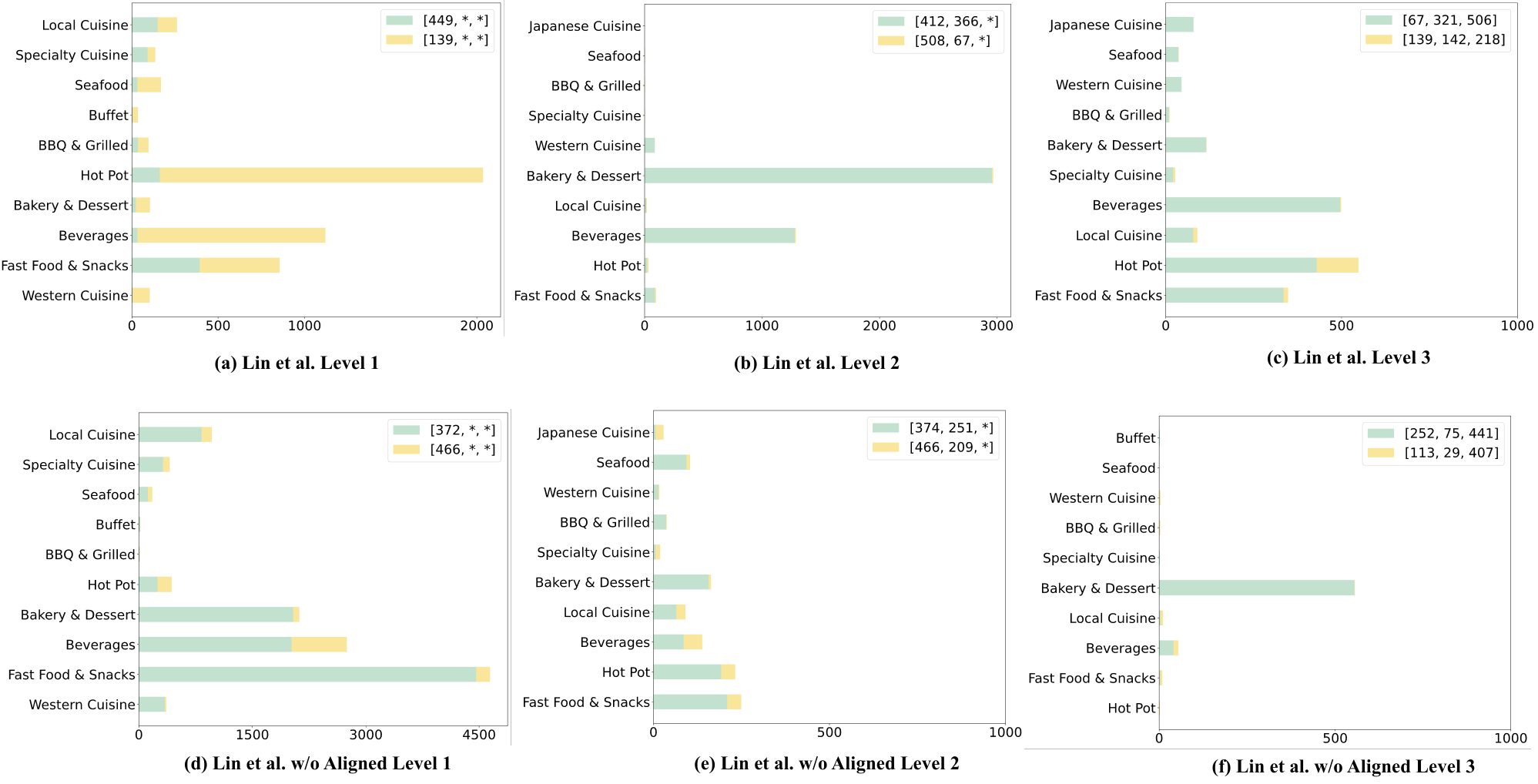}
\caption{{Hierarchical category frequency distribution of Lin et al. for different SID prefixes (Aligned vs Unaligned).}}
\label{f20}
\end{figure*}

\begin{figure*}[tp]
\includegraphics[width= \linewidth]{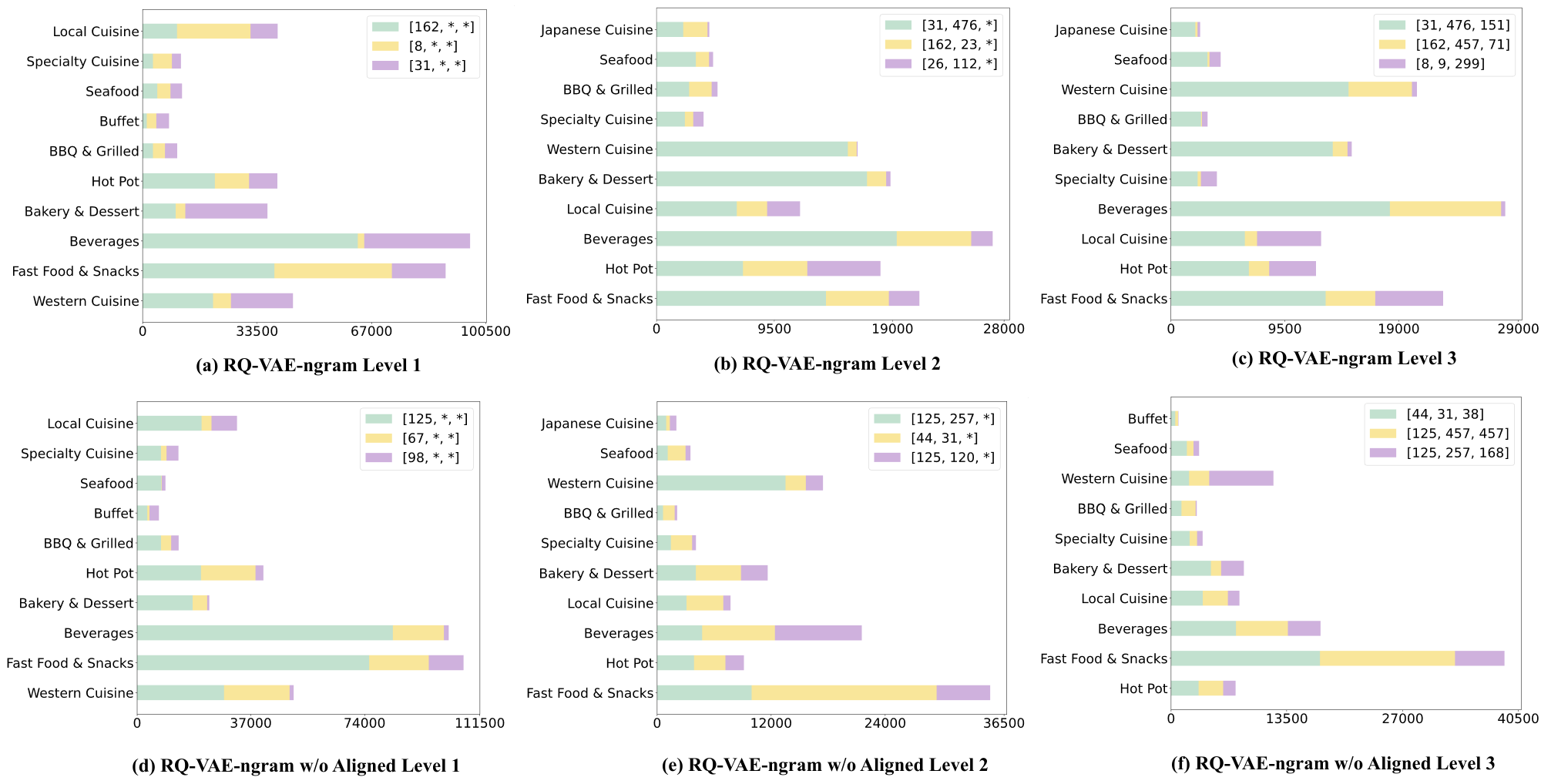}
\caption{{Hierarchical category frequency distribution of RQ-VAE-ngram for different SID prefixes (Aligned vs Unaligned).}}
\label{f21}
\end{figure*}

Figure \ref{f19}(a) depicts the Level-1 token allocation when the RQ-VAE hierarchical quantizer is trained with the G-DPO-aligned LLM embeddings.  Because this layer still leverages the same pre-computed geographic clusters as the baseline, the coarse distribution of cuisine categories—Local Cuisine, Specialty Cuisine, Seafood, Buffet, BBQ \& Grilled, Hot Pot, Bakery \& Dessert, Beverages, Fast Food \& Snacks, Western Cuisine—looks broadly similar to the un-aligned counterpart in figure \ref{f19}(d).  Crucially, however, the aligned first-layer token [162, *, *] cleanly captures the entire Local Cuisine region, while [31, *, *] does the same for Specialty Cuisine, enabling the quantizer to hand a homogeneous root to the next level; this cohesion is the direct result of the RL-based reward that encourages distance-aware grouping.

Figure \ref{f20} shows the Level-1, Level-2, and Level-3 token assignments produced by the Lin et al. hierarchical quantizer both with and without G-DPO alignment. Despite relying on the same coarse clusters, the aligned system assigns [449, *, *] to the whole Local Cuisine region and [412, *, *] to Japanese Cuisine, eliminating the scatter seen in the un-aligned case.  The single-root assignment proves that G-DPO steers the LLM to respect distance-based rewards, ensuring downstream levels receive unambiguous parent tokens.

Figure \ref{f21} depicts the Level-1 tokens for the RQ-VAE-ngram variant with G-DPO alignment.  Similar to the other aligned figures, Local Cuisine is captured by [162, *, *] and Specialty Cuisine by [31, *, *], while remaining cuisines stay compactly grouped.  The absence of fragmentation underlines that ngram-aware embeddings, once aligned, still honour geographic cohesion, providing a stable foundation for the quantizer’s deeper layers.

\section{Appendix D: Experiment}
\begin{table}[htp]
\centering
\scriptsize
\setlength{\tabcolsep}{4pt}
\begin{tabular}{l|cc cc cc}
\toprule
& \multicolumn{2}{c}{mean$\pm$SD} & \multicolumn{1}{c}{95\% CI} & \multicolumn{1}{c}{median (IQR)} \\
\cmidrule(lr){2-3} \cmidrule(lr){4-4} \cmidrule(lr){5-5}
& origin & +LGSID & +LGSID & +LGSID \\
\midrule
DIN  & $0.5859$ & $0.6259\!\pm\!.0013$* & $[0.6248,0.6269]$ & $0.6261(0.0020)$ \\
DIEN & $0.6255$ & $0.6482\!\pm\!.0019$* & $[0.6462,0.6502]$ & $0.6483(0.0022)$ \\
SIM  & $0.5884$ & $0.6259\!\pm\!.0022$* & $[0.6236,0.6282]$ & $0.6269(0.0035)$ \\
TWIN & $0.5898$ & $0.6272\!\pm\!.0022$* & $[0.6250,0.6295]$ & $0.6276(0.0030)$ \\
ETA  & $0.5903$ & $0.6272\!\pm\!.0010$* & $[0.6261,0.6283]$ & $0.6274(0.0011)$ \\
\bottomrule
\end{tabular}
\caption{Comparison of origin baseline and +LGSID on five backbones.  
     Reported are mean AUC (over 6 random seeds) with 95\% confidence interval (CI) and median (IQR).  
     * denotes that +LGSID significantly outperforms the corresponding origin baseline (Wilcoxon signed-rank, $p<0.05$).}
\label{tab:LGSID}
\end{table}

\begin{table}[t]
\centering
\begin{tabular}{l|ccc}
\toprule
\multicolumn{4}{c}{\textbf{Similarity}} \\
\cmidrule{1-4}
\textbf{$\lambda$} & Top@5 & Top@10 & Top@100 \\
\midrule
1.0 & 0.8856 & 0.8754 & 0.8283 \\
1.5 & 0.8920 & 0.8828 & 0.8395 \\
1.8 & 0.8977 & 0.8892 & 0.8504 \\
\midrule
\multicolumn{4}{c}{\textbf{Province Coverage (P@K)}} \\
\cmidrule{1-4}
\textbf{$\lambda$} & P@5 & P@10 & P@100 \\
\midrule
1.0 & 0.9960 & 0.9936 & 0.9662 \\
1.5 & 0.9943 & 0.9908 & 0.9556 \\
1.8 & 0.9905 & 0.9852 & 0.9347 \\
\midrule
\multicolumn{4}{c}{\textbf{City Coverage (C@K)}} \\
\cmidrule{1-4}
\textbf{$\lambda$} & C@5 & C@10 & C@100 \\
\midrule
1.0 & 0.9548 & 0.9352 & 0.8130 \\
1.5 & 0.9408 & 0.9152 & 0.7673 \\
1.8 & 0.9173 & 0.8858 & 0.7065 \\
\midrule
\multicolumn{4}{c}{\textbf{Town Coverage (T@K)}} \\
\cmidrule{1-4}
\textbf{$\lambda$} & T@5 & T@10 & T@100 \\
\midrule
1.0 & 0.4030 & 0.3525 & 0.2260 \\
1.5 & 0.3431 & 0.2910 & 0.1688 \\
1.8 & 0.2924 & 0.2432 & 0.1290 \\
\bottomrule
\end{tabular}
\caption{Detailed evaluation of different similarity constraint weights \(\lambda\) in G-DPO across similarity and geographic coverage metrics at Top@K levels.}
\label{tab:lambda_sensitivity}
\end{table}

\subsection{Robustness Analysis}
Table \ref{tab:LGSID} compare origin baseline and +LGSID on five backbones.  
Mean ± SD gauges the average AUC (higher is better) and its variability (lower SD means higher stability); the 95\% CI, derived from a t-distribution, narrows with lower experimental error; median and IQR measure distributional compactness, where a smaller IQR indicates insensitivity to extreme seeds; and the Wilcoxon signed-rank test compares each +LGSID column against its origin baseline with six paired seeds, deeming p $<$ 0.05 significant. 
DIN rises from 0.5859 to 0.6259 with a standard deviation of 0.0013, a 95\% confidence interval spanning only 0.0021 and an inter-quartile range of 0.0020, yielding an absolute gain of 4.0 AUC and a relative uplift of 6.8\% while also demonstrating exceptional stability under Wilcoxon testing at p equals 0.031. DIEN climbs from 0.6255 to 0.6482 with standard deviation 0.0019, confidence interval width 0.0040 and inter-quartile range 0.0022, adding 2.27 AUC and 3.6\% relative improvement while maintaining millimetre-level dispersion and confirmed steadiness at p equals 0.031. SIM jumps from 0.5884 to 0.6259 with standard deviation 0.0022 and inter-quartile range 0.0035; although seed 1 dips to 0.6224, the remaining five seeds converge around 0.627, delivering 3.75 AUC and 6.4\% relative gain while staying within low-variance bounds and proving robust at p equals 0.031. TWIN advances from 0.5898 to 0.6272 with standard deviation 0.0022, confidence interval width 0.0045 and inter-quartile range 0.0030, securing 3.74 AUC and 6.3\% relative uplift with few outliers and stable improvement confirmed at p equals 0.031. ETA ascends from 0.5903 to 0.6272 with the smallest inter-quartile range of 0.0011 and joint-lowest standard deviation of 0.0010, producing the tightest distribution, 3.69 AUC and 6.3\% relative gain with maximal reproducibility at p equals 0.031. 
Across all backbones, SDs stay small magnitude, CIs remain $\leq$ 0.005 wide, medians and means align almost perfectly, and Wilcoxon tests uniformly confirm significance, collectively evidencing that LGSID not only universally elevates performance but also endows every backbone with stability against random seeds.

\subsection{Parameters Sensitivity}
We perform extensive experiments on the similarty loss weight hyperparameter  \(\lambda\) in G-DPO. The results show that adjusting the similarity constraint allows for an effective trade-off between semantic understanding and geographic perception.
As illustrated in Table.~\ref{tab:lambda_sensitivity}, tuning the textual similarity weight \(\lambda\) enables flexible control over the LLM’s balance between content understanding and spatial perception. A higher \(\lambda\) emphasizes semantic similarity, whereas a lower value enhances geographic sensitivity. These results verify the generality of our method, enabling task-specific adaptation of the LLM by freely tuning its sensitivity to semantic and geographic signals.

\end{document}